\documentclass[11pt]{article}
\usepackage{amssymb}
\usepackage{amsfonts}

\usepackage{amsmath}

\makeatletter \@addtoreset{equation}{section} \makeatother

\newtheorem{theorem}{Theorem}
\newtheorem{lemma}{Lemma}

\newtheorem{proposition}{Proposition}

\def\tr{\mathrm{Tr}}
\def\var{\mathbf{Var}}
\def\rank{\mathrm{rank}}

\begin{document}
\pagestyle{plain}
\title{Distribution of eigenvalues of sample covariance matrices with tensor product samples}
\author{D. Tieplova}
\date{}

\setcounter{page}{1} \maketitle
\begin{abstract}
We consider $n^2\times n^2$ real symmetric and hermitian matrices $M_n$, which are equal to sum of $m_n$ tensor products of vectors $X^\mu=B(Y^\mu\otimes Y^\mu)$, $\mu=1,\dots,m_n$,
where $Y^\mu$ are i.i.d. random vectors from $\mathbb R^n (\mathbb C^n)$ with zero mean and unit variance of components, and $B$ is an $n^2\times n^2$ positive definite non-random matrix.
We prove that if $m_n/n^2\to c\in [0,+\infty)$ and the Normalized Counting Measure of eigenvalues of $BJB$, where $J$ is defined below in (\ref{J}), converges weakly, then the
Normalized Counting Measure of eigenvalues of
$M_n$ converges weakly in probability to a non-random limit and its Stieltjes transform can be found from a certain functional equation.
\end{abstract}


%

\section{Introduction }

Sample covariance matrices appeared initially in statistics in the
1920s~-1930s. Nowadays these random
matrices are widely used in statistical mechanics, probability theory and statistics, combinatorics,
operator theory and theoretical computer science
in mathematics, and also telecommunication theory, qualitative
finances, structural mechanics, etc. (see e. g. \cite{AB:11}).

We consider sample covariance matrices of the form:
\begin{align}\label{covar_ensem}
M_n=\dfrac{1}{n}XTX^*,
\end{align}
where $X$ is an $n\times m$ matrix whose entries are i.i.d. random
variables such that
\begin{align}\label{1.cond}
\mathbf E\{X_{ij}\}=0, \quad \mathbf E\{X^2_{ij}\}=1
\end{align}
and $T$ is a $m\times m$ positive definite matrix. One of the first
questions in studying of ensembles of random matrices is on their Normalized Counting
Measure of eigenvalues, which is defined by formula
\begin{align*}
N_n(\Delta) = Card\{i\in [1,n] : \lambda_i\in\Delta\}/{n},
\end{align*}
where
\begin{align*}
-\infty<\lambda_1\le \ldots \le\lambda_{n}<\infty
\end{align*}
are the eigenvalues of $M_n$.
Also let $\sigma_m$ be the Normalized Counting Measure
of eigenvalues $\{\tau_i\}_{i=1}^m$ of $T$.

First rigorous result on the model (\ref{covar_ensem}) was obtained
in \cite{MP:67}, where it was proved that if $\{m_n\}$ is a sequence of positive integers such that
\begin{align*}
m_n\rightarrow +\infty,\; n\rightarrow +\infty,\; c_n=m_n/n\rightarrow c\in[0, +\infty),
\end{align*}
and the sequence
$\sigma_m$ converges weakly to the probability measure $\sigma$:
\begin{align*}
\lim_{n \rightarrow \infty}{\sigma_m}=\sigma,
\end{align*}
 then the Normalized Counting Measure $N_n$ of eigenvalues $M_n$ converges weakly in probability to a non-random measure $N$ ($N(\mathbb{R})=1$). The Stieltjes transform $f$ of $N$,
\begin{align*}
f(z)=\int\dfrac{N(d\lambda)}{\lambda-z}, \quad\Im z \ne 0,
\end{align*}
  is uniquely determined by the equation
 \begin{align*}
f(z)=\Big(c\displaystyle{\int{\dfrac{\tau
\sigma(d\tau)}{1+\tau f(z)}-z}}\Big)^{-1}.
 \end{align*}
 Since then a lot of ensembles were considered. We mention two versions of
ensembles of sample covariance matrices, similar to
(\ref{covar_ensem}). The first is
 \begin{align}\label{si-ba}
BXX^*B,
\end{align}
 where $X$ is an $n\times m$ matrix whose entries are i.i.d. random variables satisfying (\ref{1.cond}) and $B$ is an $n\times n$ matrix. Note that while studying the
 eigenvalues of (\ref{si-ba}) we can consider the matrices $X^*B^2X$ instead of (\ref{si-ba}) coinciding with (\ref{covar_ensem}) for $T=B^2$. The second version is
\begin{align}\label{si-do}
(R_n+a X_n)(R_n+a X_n)^*,
\end{align}
where $X_n$ is an $n\times m$ matrix whose entries are i.i.d. random variables satisfying (\ref{1.cond}), $a>0$ constant, and $R_n$ is an $n\times m$ random matrix independent of $X_n$.

Numerous results and references on the eigenvalue distribution of these random matrices can be found in \cite{BS:10}, \cite{F:10}.

The paper is organized as follows. In Section 2 we present our result. In Section 3 we give the proof of the main theorem and in Section 4 we prove all the technical
results which we use in Section 3.
We denote by $C$, $c$, etc., various constants
appearing below, which can be different in different formulas.


\section{Problem and main results}\label{s:1}

Let us define multi-indexes $\mathbf{i}=(i_1, i_2)$, where $i_1, i_2=\overline{1, n}$, and inversion in  multi-indexes  $\bar{\mathbf{i}}=(i_2, i_1)$. Let
\begin{align}\label{B}
B=B_n = \{B_{\mathbf{i},\mathbf{j}}\}
\end{align}
be an $n^2\times n^2$ real symmetric or hermitian  matrix.

We consider real symmetric or hermitian random matrices
\begin{align}\label{M_n}
M_n = \dfrac{1}{n^2}\sum_{\mu=1}^m X^\mu\otimes\bar{X}^\mu,
\end{align}
where the vectors $X^\mu$ are given by the formula (cf. (\ref{si-ba}))
\begin{align}\label{X_mu}
X^\mu =  B(Y^\mu\otimes Y^\mu), \mu=1,\ldots ,m,
\end{align}
and $Y^\mu=\{Y^{\mu}_i\}_{i=1}^n$, $\mu=1,\ldots ,m$, are vectors of $\mathbb
R^n$ (or $\mathbb C^n$) such that $\{Y^{\mu}_i\}$ (or $\{\Re Y^{\mu}_i, \Im Y^{\mu}_i\}$) are i.i.d.\ random variables for all $i=\overline{1,n},\, \mu=\overline{1,m}$ and
\begin{align}\label{exp_sym}
\mathbf{E}\{Y_i^\mu\}=0,\quad \mathbf{E} \{Y_i^\mu
Y_k^\nu\}=\delta_{ik}\delta_{\mu \nu}
\end{align}
in the real symmetric case and
\begin{align}\label{exp}
\mathbf{E}\{Y_i^\mu\}=\mathbf{E} \{Y_i^\mu Y_k^\nu\}=0,\quad
\mathbf{E} \{Y_i^\mu \bar{Y}_k^\mu\}=\delta_{ik}
\end{align}
in the hermitian case.
Introduce the $n^2\times n^2$ matrix
\begin{align}\label{J}
J_{\mathbf p, \mathbf q}=\delta_{\mathbf{p}\mathbf{q}}+\delta_{\bar{\mathbf{p}} \mathbf{q}},
\end{align}
and denote by $N_n$ and $\sigma_n$ the Normalized Counting Measure of
eigenvalues of $M_n$ and $BJB$ respectively.

In what follows by saying that the matrix bounded we will mean that its euclidian (or hermitian) norm $|...|<c$ for some constant $c$. The main result of the paper is
\begin{theorem}\label{main_th}
Let $M_n$ be a random matrix defined by (\ref{B}) -- (\ref{M_n}). Assume that the sequence $\sigma_n$ converges weakly to a probability
measure $\sigma$:
\begin{align*}
\lim_{n \rightarrow \infty}{\sigma_n}=\sigma,
\end{align*}
$B$ is bounded uniformly in $n$, and $\{m_n\}$ is a sequence of positive
integers such that
\begin{align*}
m_n\rightarrow +\infty,\; n\rightarrow +\infty,\; c_n=m_n/n^2\rightarrow c\in[0, +\infty).
\end{align*}
Then the Normalized Counting Measures $N_n$ of eigenvalues of $M_n$
converge weakly in probability to a non-random
probability measure $N$, and if $f^{(0)}$
 is the Stieltjes transform of $\sigma$, then the Stieltjes transform $f$ of $N$ is uniquely determined by the equation

\begin{align*}
f(z)=f^{(0)}\left(\dfrac{z}{c-zf(z)-1}\right)(c-zf(z)-1)^{-1}
\end{align*}
 in the class of Stieltjes transforms of probability measures.
\end{theorem}


\section{Proof of the main result}\label{s:2}

We will prove the theorem for the technically simpler case of
hermitian matrices. The case of real symmetric matrices is
analogous.
Next Proposition sets the one-to-one correspondence between finite nonnegative measures
and their Stieltjes transforms.
\begin{proposition}\label{stil}
 Let $f$ be the Stieltjes transform of a finite nonnegative
measure $m$. Then:

(i)$f$ is analytic in $\mathbb C \backslash \mathbb R$, and $\overline{f(z)} = f(\overline z)$;

(ii) $\Im f(z)\Im z > 0$ for $\Im z \ne 0$;

(iii) $|f(z)| \le m(R)/|\Im z|$, in particular, $\lim\limits_{\eta\rightarrow +\infty}\eta |f(i\eta)|\le \infty$;

(iv) for any function $f$ possessing the above properties there exists a nonnegative
finite measure $m$ on $\mathbb R$ such that $f$ is its Stieltjes transform and
\begin{align}\label{mf}
\lim_{\eta\rightarrow +\infty}\eta |f(i\eta)| = m(\mathbb R);
\end{align}

(v) if $\Delta$ is an interval of $\mathbb R$ whose edges are not atoms of the measure $m$, then
we have the Stieltjes-Perron inversion formula
\begin{align*}
m(\Delta)=\lim_{\varepsilon\rightarrow +0}\frac{1}{\pi}\int\limits_{\Delta}\Im f(\lambda +i\varepsilon)d\lambda;
\end{align*}

(vi) the above one-to-one correspondence between finite nonnegative measures
and their Stieltjes transforms is continuous if we use the uniform convergence of
analytic functions on a compact set of infinite cardinality of $\mathbb C \backslash\mathbb R$ for Stieltjes
transforms and the vague convergence for measures in general
and the weak convergence of probability measures if the r.h.s. of (\ref{mf}) is 1;

\end{proposition}

 For the proofs of assertions see e.g.~\cite[Section~59]{AG:93} and~\cite{GH:03}.
Now recall some facts from linear algebra on the resolvent of real symmetric or hermitian matrix:
\begin{proposition}Let $M$ be a real symmetric (hermitian) matrix and
\begin{align*}
G_M(z)=(M-z)^{-1}, \Im z\neq 0,
\end{align*}
be its resolvent. We have:

(i) \begin{align}\label{g_bound} |G_M(z)|\leq {|\Im z|}^{-1};
\end{align}

(ii) if $G_1(z)$ and $G_2(z)$ are resolvents of real symmetric (hermitian) matrices $M_1$ and $M_2$ respectively then:
\begin{align}\label{res_id}
G_2(z)=G_1(z)-G_1(z)(M_2-M_1)G_2(z);
\end{align}

(iii) if $Y \in \mathbb R^n (\mathbb C^n)$, then
\begin{align}\label{proek}
G_{M+Y\otimes \bar{Y}}=G_{M}-\dfrac{G_M(Y\otimes \bar {Y})G_M
}{1+(G_M Y, Y)},\quad\Im z\neq 0.
\end{align}
\end{proposition}
In what follows we need
\begin{align*}
Y^{\mu(\tau)}_i=Y^{\mu}_i\mathbf{1}_{|Y^{\mu}_i|\le \tau \sqrt{n}}\, ,\quad  Y^{\mu(\tau)\circ}_i=Y^{\mu(\tau)}_i-\mathbf{E}\{Y^{\mu(\tau)}_i\}.
\end{align*}
It is easy to see that these random variables satisfy condition
\begin{align}\label{cond_1}
&\mathbf{E}\{Y^{\mu(\tau)\circ}_i\}=\mathbf{E}\{(Y^{\mu(\tau)\circ}_i)^2\}=0, \quad \mathbf{E}\{|Y^{\mu(\tau)\circ}_i|^2\}=1+o(1),\; n\rightarrow +\infty, \\
&\mathbf{E}\{|Y^{\mu(\tau)\circ}_i|^k\}\le n^{(k-2)/2}\tau^{k-2}.\label{cond_2}
\end{align}
Similarly to $X^{\mu}$ and $M_n$ we can define
\begin{align*}
X^{\mu(\tau)}=B(Y^{\mu(\tau)\circ}\otimes Y^{\mu(\tau)\circ}), \quad M^{\tau}_n=\dfrac{1}{n^2}\sum_{\mu=1}^{m}X^{\mu(\tau)}\otimes \bar{X}^{\mu(\tau)}.
\end{align*}
Consider $n^2\times n^2$ matrices
\begin{align*}
K_n=\dfrac{1}{n^2}\sum_{\mu=1}^mC^{\mu}\otimes\bar{C}^{\mu}, \quad
\widehat{K}_n=\dfrac{1}{n^2}\sum_{\mu=1}^mC^{\mu}\otimes\bar{X}^{\mu},
\end{align*}
where
\begin{align}\label{def_C}
C^{\mu}_{\mathbf{i}}=\sum_{\mathbf{p}}B_{\mathbf{i},\mathbf{p}}(Y^{\mu}_{p_1}Y^{\mu}_{p_2}(1-\delta_{p_1,p_2})+Y^{\mu(\tau)\circ}_{p_1}Y^{\mu(\tau)\circ}_{p_2}\delta_{p_1,p_2}).
\end{align}
Here and below $\displaystyle\sum\limits_{\mathbf{p}}=\sum\limits_{p_1=1}^n\sum\limits_{p_2=1}^n.$

We need the following simple fact, a version of the min-max principle of linear algebra (see e. g. \cite{K:76}, Section I.6.10).
\begin{proposition}
 Let $M_1$ and $M_2$ be $n\times n$ hermitian matrices and $N_1$ and $N_2$ be Normalized
Counting Measures of their eigenvalues. Then we have for any interval $\Delta\subset \mathbb{R}$:
\begin{align}\label{min-max}
|N_1(\Delta)-N_2(\Delta)|\le \rank(A_1-A_2)/n.
\end{align}
\end{proposition}

Let $N_n$,  $N^{(1)}_n$ and $\widehat{N}^{(1)}_n$ be the Normalized Counting Measure of eigenvalues of matrices $M_n$, $K_n$ and $\widehat{K}_n$ respectively. Then according to (\ref{min-max})
and (\ref{def_C})
\begin{align*}
&|N_n-N^{(1)}_n|\le|N_n-\widehat{N}^{(1)}_n|+|\widehat{N}^{(1)}_n-N^{(1)}_n|\le\rank(M_n-\widehat{K}_n)/n^2+\rank(\widehat{K}_n-K_n)/n^2\\
&\le \dfrac{1}{n^2}\Big(\rank\{\sum_{\mathbf{p}}B_{\mathbf{i},\mathbf{p}}\{\sum^{m}_{\mu=1}(Y^{\mu(\tau)\circ}_{p_1}Y^{\mu(\tau)\circ}_{p_2}
-Y^{\mu}_{p_1}Y^{\mu}_{p_2})\delta_{p_1,p_2}\bar{X}^{\mu}_{\mathbf{q}}\}_{\mathbf{p},\mathbf{q}}\}_{\mathbf{i},\mathbf{q}}\\
&\hskip5cm+ \rank\{\sum_{\mathbf{q}}\{\sum^{m}_{\mu=1}C^{\mu}_{\mathbf{p}}(\bar{Y}^{\mu(\tau)\circ}_{q_1}\bar{Y}^{\mu(\tau)\circ}_{q_2}
-\bar{Y}^{\mu}_{q_1}\bar{Y}^{\mu}_{q_2})\delta_{q_1,q_2}\}_{\mathbf{p},\mathbf{q}}\bar{B}_{\mathbf{q},\mathbf{i}}\}_{\mathbf{p},\mathbf{i}}\Big)\\
&\le \dfrac{1}{n^2}\Big(\rank\{\sum^{m}_{\mu=1}(Y^{\mu(\tau)\circ}_{p_1}Y^{\mu(\tau)\circ}_{p_2}-
Y^{\mu}_{p_1}Y^{\mu}_{p_2})\delta_{p_1,p_2}\bar{X}^{\mu}_{\mathbf{q}}\}_{\mathbf{p},\mathbf{q}}\\
&\hskip6cm+ \rank\{\sum^{m}_{\mu=1}C^{\mu}_{\mathbf{p}}(\bar{Y}^{\mu(\tau)\circ}_{q_1}\bar{Y}^{\mu(\tau)\circ}_{q_2}-
\bar{Y}^{\mu}_{q_1}\bar{Y}^{\mu}_{q_2})\delta_{q_1,q_2}\}_{\mathbf{p},\mathbf{q}}\Big)
 =\dfrac{2}{n}.
\end{align*}
\begin{lemma}\label{l(g-gm)}
Let $G^{(1)}(z)$ and $G^{\tau}(z)$ be the resolvents of the matrices $K_n$ and $M^{\tau}_n$ respectively. Then
\begin{align*}
\dfrac{1}{n^2}|\mathbf{E}\{\tr(G^{(1)}(z)-G^{\tau}(z))\}|=o(1),\; n\rightarrow +\infty.
\end{align*}
\end{lemma}
\textit{Proof.}
Consider the $(n^2+m)\times(n^2+m)$ block matrices $\widetilde{M}_n$ and $\widetilde{M}^{\tau}_n$ such that:
\begin{align}\label{M_til}
\widetilde{M}_n=\begin{pmatrix}
0&A^{\ast}\\
A&0
\end{pmatrix}, \quad \widetilde{M}_n^{\tau}=\begin{pmatrix}
0&(A^{\tau})^{\ast}\\
A^{\tau}&0
\end{pmatrix},
\end{align}
where $A$, $A^{\tau}$ are $n^2\times m$ matrices and
\begin{align*}
A_{\mathbf{i}, \mu}=n^{-1}C^{\mu}_{\mathbf{i}},\quad A^{\tau}_{\mathbf{i}, \mu}=n^{-1}X^{\mu(\tau)}_{\mathbf{i}}.
\end{align*}
Denote $\widetilde{G}(z)$ and $\widetilde{G}^{\tau}(z)$ the resolvents of matrices $\widetilde{M}_n$ and $\widetilde{M}^{\tau}_n$ respectively.
Using formula of inversion of block matrix, we get:
\begin{align}\label{g_tild}
\tr(G^{(1)}(z^2)-G^{\tau}(z^2))
=-\dfrac{z}{2}\tr(\widetilde{G}(z)-\widetilde{G}^{\tau}(z)).
\end{align}
Now we should estimate the last expression. From (\ref{res_id}) we have:
\begin{align*}
|\tr(\widetilde{G}-\widetilde{G}^{\tau})|=|\tr(\widetilde{G}\widetilde{G}^{\tau}&(\widetilde{M}_n-\widetilde{M}_n^{\tau}))|\\
&\le(\tr(\widetilde{G}\widetilde{G}^{\tau}\widetilde{G}^{\ast}\widetilde{G}^{\tau\ast}))^{1/2}(\tr(\widetilde{M}_n-\widetilde{M}_n^{\tau})
(\widetilde{M}^{\ast}_n-\widetilde{M}_n^{\tau\ast}))^{1/2}.
\end{align*}
Here and below we drop the argument $z$.
Relations (\ref{g_bound}) and (\ref{M_til}) implies:
\begin{align*}
&|\tr(\widetilde{G}-\widetilde{G}^{\tau})|\le \dfrac{n}{\Im z^2}(\tr(2(A-A^{\tau})(A^{\ast}-(A^{\tau})^{\ast})))^{1/2}\\
&\hskip6cm=\dfrac{1}{n\Im z^2}\Big(2\sum_{\mu=1}^m\sum_{\mathbf{i}}(C^{\mu}_{\mathbf{i}}-X^{\mu(\tau)}_{\mathbf{i}})(\bar{C}^{\mu}_{\mathbf{i}}-
\bar{X}^{\mu(\tau)}_{\mathbf{i}})\Big)^{1/2}\\
&=\dfrac{n}{\Im z^2}\Big(2\sum_{\mu=1}^m\sum_{\mathbf{i},\mathbf{p},\mathbf{q}}B_{\mathbf{i},\mathbf{p}}(1-\delta_{p_1,p_2})(Y^{\mu}_{p_1}Y^{\mu}_{p_2}-
Y^{\mu(\tau)\circ}_{p_1}Y^{\mu(\tau)\circ}_{p_2})\\
&\hskip7cm\times B_{\mathbf{q},\mathbf{i}}(1-\delta_{q_1,q_2})(\bar{Y}^{\mu}_{q_1}\bar{Y}^{\mu}_{q_2}-
\bar{Y}^{\mu(\tau)\circ}_{q_1}\bar{Y}^{\mu(\tau)\circ}_{q_2})\Big)^{1/2}\\
&=\dfrac{1}{\Im z^2}\Big(2\sum_{\mu=1}^m\sum_{\substack{p_1\ne p_2\\q_1\ne q_2}}B^2_{\mathbf{q},\mathbf{p}}(Y^{\mu}_{p_1}Y^{\mu}_{p_2}\bar{Y}^{\mu}_{q_1}\bar{Y}^{\mu}_{q_2}-
 Y^{\mu(\tau)\circ}_{p_1}Y^{\mu(\tau)\circ}_{p_2}\bar{Y}^{\mu}_{q_1}\bar{Y}^{\mu}_{q_2}\\
&\hskip5cm-Y^{\mu}_{p_1}Y^{\mu}_{p_2}\bar{Y}^{\mu(\tau)\circ}_{q_1}\bar{Y}^{\mu(\tau)\circ}_{q_2}+
Y^{\mu(\tau)\circ}_{p_1}Y^{\mu(\tau)\circ}_{p_2}\bar{Y}^{\mu(\tau)\circ}_{q_1}\bar{Y}^{\mu(\tau)\circ}_{q_2})\Big)^{1/2}.
\end{align*}
Notice that in view of (\ref{cond_1}) and (\ref{exp}) entries where one of indexes $\{p_1,p_2,q_1,q_2\}$ is different from others equal zero.
Thus
\begin{align*}
&|\tr(\widetilde{G}-\widetilde{G}^{\tau})|\le\dfrac{1}{\Im z^2}\Big(2\sum_{\mu=1}^m\sum_{\substack{\mathbf{p}=\mathbf{q}\\\bar{\mathbf{p}}=\mathbf{q}}}
B^2_{\mathbf{q},\mathbf{p}}(Y^{\mu}_{p_1}Y^{\mu}_{p_2}\bar{Y}^{\mu}_{q_1}\bar{Y}^{\mu}_{q_2}- Y^{\mu(\tau)\circ}_{p_1}Y^{\mu(\tau)\circ}_{p_2}\bar{Y}^{\mu}_{q_1}\bar{Y}^{\mu}_{q_2}\\
&\hskip5cm-Y^{\mu}_{p_1}Y^{\mu}_{p_2}\bar{Y}^{\mu(\tau)\circ}_{q_1}\bar{Y}^{\mu(\tau)\circ}_{q_2}+
Y^{\mu(\tau)\circ}_{p_1}Y^{\mu(\tau)\circ}_{p_2}\bar{Y}^{\mu(\tau)\circ}_{q_1}\bar{Y}^{\mu(\tau)\circ}_{q_2})\Big)^{1/2}.
\end{align*}
Relations (\ref{cond_1}) and (\ref{exp}) implies
\begin{align*}
&\mathbf{E}\{|Y^{\mu}_{p_1}|^2|Y^{\mu}_{p_2}|^2-Y^{\mu(\tau)\circ}_{p_1}Y^{\mu(\tau)\circ}_{p_2}\bar{Y}^{\mu}_{p_1}\bar{Y}^{\mu}_{p_2}-
Y^{\mu}_{p_1}Y^{\mu}_{p_2}\bar{Y}^{\mu(\tau)\circ}_{p_1}\bar{Y}^{\mu(\tau)\circ}_{p_2}+|Y^{\mu(\tau)\circ}_{p_1}|^2|Y^{\mu(\tau)\circ}_{p_2}|^2\}\\
&=1-(1+o(1))-(1+o(1))+(1+o(1))=o(1).
\end{align*}
Combining all above we get
\begin{align*}
\dfrac{1}{n^2}|\mathbf{E}\{\tr(\widetilde{G}-\widetilde{G}^{\tau})\}|< \dfrac{(2m\tr(JB)^2o(1))^{1/2}}{N\Im z^2}= \dfrac{\sqrt{2m}}{n\Im z^2}o(1).
\end{align*}
Finally in view of (\ref{g_tild})
\begin{align*}
\dfrac{1}{n^2}|\mathbf{E}\{\tr(G(z)^{(1)}-G^{\tau}(z))\}|<\dfrac{\sqrt{m}}{\sqrt{2}n|\Im z|}o(1)=o(1).
\end{align*}

$\square$

It follows from Lemma \ref{l(g-gm)} that for our purposes it suffices to prove Theorem \ref{main_th} for matrix $M^{\tau}_n$.
Hence below we will assume that $M_n$ is replaced by $M^{\tau}_n$. To simplify notations we drop the index $\tau$ and denote
\begin{align*}
G(z)=(M_n-z)^{-1},\; G^\mu(z)=G\mid_{X^\mu=0}, \; N=n^2.
\end{align*}
In the proof of main theorem we need some results
\begin{lemma}\label{l2}
If $F$ is a non-random $N\times N$ matrix such that $|F|\leq c$
then

(i)\begin{align}\label{fgbb}
\begin{split}
&\mathbf{E}\{(FG^\mu X^\mu, X^\mu)\}=\tr(FG^\mu BJB),\\
&\var\{N^{-1}(FG^\mu X^\mu, X^\mu)\}= o(1), \; n\rightarrow +\infty;
\end{split}
\end{align}

(ii)
\begin{align}\label{g-gm}
\dfrac{1}{N}|\tr{F(G-G^\mu)}|=O(N^{-1});
\end{align}

(iii)
\begin{align}\label{tr(fg)}
\var\{N^{-1}\tr(FG)\}\leq \dfrac{c}{N}.
\end{align}
\end{lemma}
The proof of the lemma is given in Section \ref{s:3}.

According to (\ref{proek}), we have
\begin{align*}
G_{\mathbf{i}, \mathbf{j}}=G^\mu_{\mathbf{i},
\mathbf{j}}-N^{-1}\dfrac{(G^\mu X^\mu)_{\mathbf{i}}(G^\mu\bar{X}^\mu)_{\mathbf{j}}}{1+N^{-1}(G^\mu X^\mu, X^\mu)}.
\end{align*}
Hence,
\begin{align*}
(G X^\mu)_{\mathbf{i}}=\dfrac{(G^\mu X^\mu)_{\mathbf{i}}}{1+N^{-1}(G^\mu X^\mu,X^\mu)}.
\end{align*}
Take any $N\times N$ bounded matrix $K$. Then
\begin{align}\label{tr}
&\dfrac{1}{N}\tr(KGM)=\dfrac{1}{N^2}\sum_{\mu=1}^{m}\sum_{\mathbf i, \mathbf j}K_{\mathbf j, \mathbf i}(GX^\mu)_{\mathbf i}\bar{ X}^\mu_{\mathbf{j}}\notag\\
&=\dfrac{1}{N^2}\sum_{\mu=1}^{m}\sum_{\mathbf{j}}\dfrac{(KG^\mu
X^\mu)_{\mathbf{j}}\bar{X}^\mu_{\mathbf{j}}}{1+N^{-1}(G^\mu X^\mu,
X^\mu)}= \dfrac{1}{N^2}\sum_{\mu=1}^{m}\dfrac{(KG^\mu X^\mu,
X^\mu)}{1+N^{-1}(G^\mu X^\mu, X^\mu)}.
\end{align}

To analyze the r.h.s. of (\ref{tr}), let us show first that if
$\mathcal C$ and $\mathcal D$ are random variables, such that
$\mathbf E\{|\mathcal C|^2+|\mathcal D|^2\}< c$ and
\begin{equation*}
\bar{\mathcal C}=\mathbf{E}\{\mathcal C\},\quad \mathcal
{C^{\circ}}=\mathcal C-\bar{\mathcal C},\quad\bar{\mathcal
D}=\mathbf{E}\{\mathcal D\}, \quad \mathcal D^{\circ}=\mathcal
D-\bar{\mathcal D},
\end{equation*}
then
\begin{align}\label{exp_frac}
\mathbf{E}\left\{\dfrac{\mathcal C}{\mathcal
D}\right\}=\dfrac{\bar{\mathcal C}}{\bar{\mathcal
D}}+O\left(\mathbf{E}\left\{\dfrac{|\mathcal
C^\circ|^2}{|\bar{\mathcal D}|^2}+ \dfrac{|\mathcal
D^\circ|^2}{|\bar{\mathcal D}|^2}\right\}\right).
\end{align}
Indeed,
\begin{align*}
\dfrac{\mathcal C}{\mathcal D}=\dfrac{\bar{\mathcal C}+\mathcal
C^{\circ}}{\bar{\mathcal D}}-\dfrac{(\bar{\mathcal C}+\mathcal
C^{\circ})\mathcal D^{\circ}}{\bar{\mathcal D}^2}+
O\left(\left(\dfrac{\mathcal D^\circ}{\bar{\mathcal
D}}\right)^3\right).
\end{align*}
Thus
\begin{align*}
\mathbf{E}\left\{\dfrac{\mathcal C}{\mathcal
D}\right\}=\dfrac{\bar{\mathcal C}}{\bar{\mathcal
D}}+\mathbf{E}\left\{\dfrac{\mathcal C^{\circ}\mathcal
D^{\circ}}{\bar{\mathcal D}^2}
\right\}+O\left(\dfrac{|\mathcal D^{\circ}|^3}{\bar{\mathcal D}^3}\right) \leq
\dfrac{\bar{\mathcal C}}{\bar{\mathcal
D}}+\mathbf{E}\left\{\dfrac{|\mathcal C^\circ|^2}{|\bar{\mathcal
D}|^2}+c\dfrac{|\mathcal D^\circ|^2}{|\bar{\mathcal
D}|^2}\right\}.
\end{align*}
The last inequality implies (\ref{exp_frac}).

Let $\mathcal C=N^{-1}(KG^\mu X^\mu,X^\mu)$, $\mathcal
D=1+2N^{-1}(G^\mu X^\mu,X^\mu)$. Since matrix $K$ is bounded, it follows from (\ref{fgbb}) that
\begin{align*}
\mathbf E_{\mu}\{|\mathcal C^{\circ}|^2\}=\mathbf E_{\mu}\{|\mathcal D^{\circ}|^2\}=o(1),\,n\rightarrow +\infty.
\end{align*}
 This,  (\ref{tr}) and (\ref{exp_frac}) imply
\begin{align}\label{kgm}
\dfrac{1}{N}\mathbf E\{\tr(KGM)\}=\dfrac{1}{N}\sum_{\mu=1}^m
\Big(\mathbf
E\Big\{\dfrac{N^{-1}\tr(KG^{\mu}BJB)}{1+N^{-1}\tr(G^{\mu}BJB)}\Big\}+o(1)\Big).
\end{align}
In the r.h.s. of (\ref{kgm}) result (\ref{g-gm}) allows us to replace $G^\mu$
with $G$
\begin{align}\label{kgm2}
\dfrac{1}{N}\mathbf{E}\{\tr(KGM)\}=\mathbf{E}\Big\{\dfrac{c_n N^{-1}\tr(KGBJB)}{1+N^{-1}\tr(GBJB)}+o(1))\Big\}.
\end{align}
The last step is to replace $N^{-1}\tr (KGBJB)$ and
$N^{-1}\tr (GBJB)$ in (\ref{kgm2}) with their expectations. We use again (\ref{exp_frac}) with $\mathcal
C=N^{-1}\tr(KGBJB)$, $\mathcal D=1+N^{-1}\tr(GBJB)$.
It follows from (\ref{kgm2}) and (\ref{tr(fg)})
\begin{align}\label{kgm_ex}
\dfrac{1}{N}\mathbf{E}\{\tr(KGM)\}=\dfrac{c_n N^{-1}\mathbf{E}\{\tr(KGBJB)\}}{1+N^{-1}\mathbf{E}\{\tr(GBJB)\}}+o(1).
\end{align}
Note that
\begin{align*}
\dfrac{1}{N}\mathbf{E}\{\tr(KGM)\}=\dfrac{1}{N}\mathbf{E}\{\tr(K(G(M-z)+Gz))\}=\dfrac{1}{N}\mathbf{E}\{\tr K\}+
\dfrac{z}{N}\mathbf{E}\{\tr(KG)\}.
\end{align*}
This and (\ref{kgm_ex}) imply that for
any bounded matrix $K$
\begin{align} \label{kfin}
\dfrac{1}{N}\mathbf{E}\{\tr
K\}=\dfrac{1}{N}\mathbf{E}\{\tr(KG(c_n
b^{-1}_nBJB-z))\}+o(1),
\end{align}
where
\begin{align}\label{b_n}
b_n=1+N^{-1}\mathbf{E}\{\tr(GBJB)\}.
\end{align}
Taking $K=(c_n b^{-1}_nBJB-z)^{-1}$, we obtain
\begin{align}\label{f_n}
\dfrac{1}{N}\mathbf{E}\{\tr(c_n
b^{-1}_nBJB-z)^{-1}\}=f_n(z)+o(1),
\end{align}
where
\begin{align*}
g_n(z)=\dfrac{1}{N}\tr(G(z)),\quad f_n(z)=\mathbf{E}\{g_n(z)\}.
\end{align*}
It follows from (\ref{kfin}) with $K=I$
\begin{align*}
\dfrac{1}{N}\mathbf{E}\{\tr
(I+zG)\}=\dfrac{c_n}{b_n}(b_n-1)+o(1).
\end{align*}
Then we get
\begin{align*}
1+zf_n(z)=c_n(1-\dfrac{1}{b_n})+o(1).
\end{align*}
Now we can find $b_n$:
\begin{align}\label{c_n/b_n}
b_n=\dfrac{c_n}{c_n-zf_n(z)-1+o(1)}.
\end{align}
This and (\ref{f_n}) yield
\begin{align}\label{f_n_fin}
f_n(z)=f_n^{(0)}\left(\dfrac{z}{c_n-zf_n(z)-1}\right)({c_n-zf_n(z)-1})^{-1}+o(1),
\end{align}
where
\begin{align*}
f^{(0)}_n(z)=\dfrac{1}{N}\mathbf{E}\{\tr(BJB-z)^{-1}\}.
\end{align*}
The sequence $\{f_n\}$ consists of functions, analytic and uniformly bounded in
$n$ and~$z$. Hence, there exists an analytic in $\mathbb C\backslash \mathbb R$ function $f$ and a
subsequence $\{f_{n_j}\}$ that converges to $f$ uniformly on any compact set of $\mathbb C\backslash\mathbb R$.
In addition we have
\begin{align*}
\Im f_n(z)\Im z> 0,\; \Im z\ne 0
\end{align*}
thus $\Im f(z)\Im z\ge 0,\; \Im z\ne 0$. By Proposition~\ref{stil}(vi) and the hypothesis of the
theorem on the weak convergence of the sequence $\sigma_n$ to $\sigma$, the sequence
$f_n^{(0)}$ of their Stieltjes transforms consists of analytic in $\mathbb C\backslash \mathbb R$ functions that
converge uniformly on a compact set of $\mathbb C\backslash \mathbb R$ to the Stieltjes transform $f^{(0)}$ of
the limiting counting measure $\sigma$ of matrices $BJB$. This allows
us to pass to the limit $n\rightarrow +\infty$ in (\ref{f_n_fin}) and to obtain that the limit $f$ of any
converging subsequence of the sequence $f_n$ satisfies functional equation
\begin{align}\label{f_fin}
f(z)=f^{(0)}\left(\dfrac{z}{c-zf(z)-1}\right)\Big(c-zf(z)-1\Big)^{-1},
\end{align}
and $\Im f(z)\Im z\ge 0,\; \Im z\ne 0$. The proof of the uniqueness of solution of the equation in the class of functions, analytic for $\Im z \ne 0$ and such that $\Im f(z)\Im z\ge 0,\; \Im z\ne 0$ is
analogues to \cite{MP:67}. Hence, the whole sequence $f_n$ converges uniformly
on a compact set of $\mathbb C\backslash \mathbb R$ to the unique solution $f$ of the equation. Let's show that the solution possesses the properties $\Im f(z)\Im z\ge 0,\; \Im z\ne 0$ and $\lim\limits_{\eta\rightarrow +\infty}\eta |f(i\eta)| =1$.
Assume that $\Im f(z_0)=0,\; \Im z_0\ne 0$. Then~(\ref{f_fin}) implies that
\begin{align*}
\Im \int\dfrac{d\sigma (\lambda)}{(c-1)\lambda-z_0(f(z_0)-1)}=C\Im f^{(0)}(\tilde{z})=0,
\end{align*}
where $C$ is some real constant and $\Im \tilde{z}\ne 0$. This is impossible because, according to Proposition~\ref{stil}(ii), $\Im f^{(0)}(z)$ is strictly positive for any nonreal $z$. Since $|f(i\eta)|<\eta^{-1}$ we have
\begin{align*}
\lim_{\eta\rightarrow +\infty}\eta |f(i\eta)| =\lim_{\eta\rightarrow +\infty}\int\dfrac{\eta d\sigma (\lambda)}{(c-1)\lambda-i\eta -i\eta f(i\eta)}=1
\end{align*}
This and the Proposition~\ref{stil}(iv) imply that $f$ is Stieltjes
transform of a probability measure.

$ \square$

\section{Proofs of the lemma ~\ref{l2}}\label{s:3}


$(i)$ It follows from (\ref{exp})
\begin{align*}
\mathbf E_{\mu}\{(FG^\mu X^\mu, X^\mu)\}=\tr(FG^{\mu}BJB).
\end{align*}
Denote
\begin{align*}
r_n^\mu=(FG^\mu X^\mu, X^\mu)-\tr(FG^{\mu}BJB).
\end{align*}
We need to show that $\mathbf E_{\mu}\{(N^{-1}r^\mu)^2\}=o(1),\, n\rightarrow +\infty$.
Rewrite
\begin{align*}
r_n^{\mu}&=\sum_{\mathbf i, \mathbf j, \mathbf p, \mathbf q}(FG^\mu)_{\mathbf i, \mathbf j}B_{\mathbf j, \mathbf p}B_{\mathbf q, \mathbf i}
(Y^\mu_{p_1}Y^\mu_{p_2}\bar {Y}^\mu_{q_1}\bar {Y}^\mu_{q_2}-J_{\mathbf p, \mathbf q})\\
&=\sum_{\mathbf i, \mathbf j}(FG^\mu)_{\mathbf i, \mathbf j}
\Big(\sum_{\mathbf p}B_{\mathbf j, \mathbf p}B_{\mathbf p, \mathbf i}\Big(|Y^\mu_{p_1}|^2|Y^\mu_{p_2}|^2-1\Big)\\
&\hskip2cm+ \sum_{\mathbf p}B_{\mathbf j, \mathbf p}B_{\mathbf{\bar p}, \mathbf i}\Big(|Y^\mu_{p_1}|^2|Y^\mu_{p_2}|^2-1\Big)
+ \sum_{\substack{\mathbf p\neq\mathbf q\\\bar{\mathbf p}\neq \mathbf q}}B_{\mathbf j, \mathbf p}Y^\mu_{p_1}Y^\mu_{p_2}
B_{\mathbf q, \mathbf i}\bar {Y}^\mu_{q_1}\bar {Y}^\mu_{q_2}\Big)\\
&=\sum_{\mathbf i, \mathbf j}(FG^\mu)_{\mathbf i,
\mathbf j}
\Big(\sum_{\mathbf p}B_{\mathbf j, \mathbf p}(JB)_{\mathbf p, \mathbf i}\Big(|Y^\mu_{p_1}|^2|Y^\mu_{p_2}|^2-1\Big)\\
&\hskip8cm +\sum_{\substack{\mathbf p\neq\mathbf q\\\bar{\mathbf
p}\neq \mathbf q}}B_{\mathbf j, \mathbf p}Y^\mu_{p_1}Y^\mu_{p_2}
B_{\mathbf q, \mathbf i}\bar {Y}^\mu_{q_1}\bar {Y}^\mu_{q_2}\Big).
\end{align*}
Since $G^\mu$ is independent of $Y^\mu$, we obtain
\begin{align*}
&\mathbf E_{\mu}\{(N^{-1}r^\mu)^2\}=\dfrac{1}{N^2}\mathbf
E_{\mu}\Big\{\Big(\sum_{\mathbf i, \mathbf j}(FG^\mu)_{\mathbf i,
\mathbf j} \Big)^2
\Big(\sum_{\mathbf p}B_{\mathbf j, \mathbf p}(JB)_{\mathbf p, \mathbf i}\Big(|Y^\mu_{p_1}|^2|Y^\mu_{p_2}|^2-1\Big)\\
&\hskip10cm +\sum_{\substack{\mathbf p\neq\mathbf q\\\bar{\mathbf
p}\neq \mathbf q}}B_{\mathbf j, \mathbf p}Y^\mu_{p_1}Y^\mu_{p_2}
B_{\mathbf q, \mathbf i}\bar {Y}^\mu_{q_1}\bar {Y}^\mu_{q_2}\Big)^2\Big\}\\
&=\dfrac{1}{N^2}\mathbf E_{\mu}\Big\{\sum  _{\mathbf i, \mathbf
j}\sum_{\mathbf i', \mathbf j'}(FG^\mu)_{\mathbf i, \mathbf
j}(\bar{F}\bar{G}^\mu)_{\mathbf i', \mathbf j'}
\Big( \sum_{\substack{\mathbf p\neq\mathbf q\\\bar{\mathbf p}\neq \mathbf q}}\sum_{\substack{\mathbf p'\neq\mathbf q'\\\bar{\mathbf p}'\neq \mathbf q'}}
B_{\mathbf j, \mathbf p}Y^\mu_{p_1}Y^\mu_{p_2}B_{\mathbf q,
\mathbf i}\bar {Y}^\mu_{q_1}\bar {Y}^\mu_{q_2}\bar {B}_{\mathbf
j', \mathbf p'}\bar{Y}^\mu_{p'_1}\bar{Y}^\mu_{p'_2} \bar
{B}_{\mathbf q', \mathbf i'}Y^\mu_{q'_1}Y^\mu_{q'_2}\Big\}
\\
&\hskip9,5cm+\dfrac{1}{N^2}\mathbf E_{\mu}\Big\{\sum_{\mathbf i, \mathbf j}\sum_{\mathbf i', \mathbf j'}(FG^\mu)_{\mathbf i, \mathbf j}\bar{(FG^\mu)}_{\mathbf i', \mathbf j'}\\
&\times\sum_{\mathbf p}\sum_{\mathbf p'}B_{\mathbf j, \mathbf
p}(JB)_{\mathbf p, \mathbf i}\bar{B}_{\mathbf j', \mathbf
p'}(J\bar{B})_{\mathbf p', \mathbf i'}
\Big(|Y^\mu_{p_1}|^2|Y^\mu_{p_2}|^2-1\Big)\Big(|Y^\mu_{p'_1}|^2|Y^\mu_{p'_2}|^2-1\Big)\Big\}\\
&\hskip9,5cm+\dfrac{2}{N^2}\mathbf E_{\mu}\Big\{\sum_{\mathbf i, \mathbf j}\sum_{\mathbf i', \mathbf j'}(FG^\mu)_{\mathbf i, \mathbf j}\bar{(FG^\mu)}_{\mathbf i', \mathbf j'}\\
&\times\sum_{\mathbf p}\sum_{\substack{\mathbf p'\neq\mathbf
q'\\\bar{\mathbf p}'\neq \mathbf q'}}B_{\mathbf j, \mathbf
p}(JB)_{\mathbf p, \mathbf i}
\Big(|Y^\mu_{p_1}|^2|Y^\mu_{p_2}|^2-1\Big)\bar {B}_{\mathbf j',
\mathbf p'}\bar{Y}^\mu_{p'_1}\bar{Y}^\mu_{p'_2} \bar {B}_{\mathbf
q', \mathbf
i'}Y^\mu_{q'_1}Y^\mu_{q'_2}\Big)\Big\}=:\dfrac{1}{N^2}(R_1+R_2+R_3).
\end{align*}
Denote
\begin{align*}
H=BFG^\mu B,
\end{align*}
and introduce an $N\times N$ matrix $\Delta$ such that
\begin{align*}
\Delta_{\mathbf i, \mathbf j}=\delta_{i_1 j_2}\delta_{i_2 j_1}.
\end{align*}
It is easy to check that for any $N\times N$ matrix $A$
\begin{align}\label{A_inv}
\begin{split}
&A_{i_2i_1, j_1j_2}=(\Delta A)_{\mathbf i, \mathbf j},\\
&A_{i_1i_2, j_2j_1}=(A\Delta)_{\mathbf i, \mathbf j}.
\end{split}
\end{align}
Let us define the set $E=\{p_1, p_2, q_1, q_2, p'_1, p'_2, q'_1, q'_2\}$. Note that if in the set E more then 4 different numbers that
\begin{align*}
\mathbf{E}_{\mu}\{Y^\mu_{p_1}Y^\mu_{p_2}\bar {Y}^\mu_{q_1}\bar {Y}^\mu_{q_2}\bar{Y}^\mu_{p'_1}\bar{Y}^\mu_{p'_2} Y^\mu_{q'_1}Y^\mu_{q'_2}\}=0.
\end{align*}
Hence we need to consider the sets  $I_1$, $I_2$, $I_3$ and $I_4$ of all multi-indexes $\{\mathbf p, \mathbf
q, \mathbf p', \mathbf q'\}$ of the special form:
\begin{align*}
I_1=\Big\{\{\mathbf p, \mathbf q, \mathbf p', \mathbf q'\}=\{(a, b), (a,c), (d, b), (d, c)\}\Big\},\\
I_2=\Big\{\{\mathbf p, \mathbf q, \mathbf p', \mathbf q'\}=\{(a, b), (c,d), (a, b), (c, d)\}\Big\},
\end{align*}
where numbers $a, b, c$ and $d$ are all pairwise different,
\begin{align*}
I_3=\Big\{\{\mathbf p, \mathbf q, \mathbf p', \mathbf q'\}: \text{there are 3 different numbers (i, j, k) in the set } E \Big\},\\
I_4=\Big\{\{\mathbf p, \mathbf q, \mathbf p', \mathbf q'\}: \text{there are 2 different numbers (i, j) in the set } E \Big\}
\end{align*}
or any inversion in the multi-indexes of such form. Since $B$, $F$, $\Delta$ and $G^\mu$ (in view of (\ref{g_bound})) are
bounded, then there exists a constant $c$ such that $|H|<c$. Hence in view of (\ref{A_inv}) and (\ref{cond_2})
\begin{align*}
R_1&\le\mathbf E_{\mu}\Big\{\sum_{I_1}H_{\mathbf p, \mathbf q}\bar{H}_{\mathbf p', \mathbf q'}|Y^{\mu}_a|^2|Y^{\mu}_b|^2|Y^{\mu}_c|^2|Y^{\mu}_d|^2+
\sum_{I_2}H_{\mathbf p, \mathbf q}\bar{H}_{\mathbf p', \mathbf q'}|Y^{\mu}_a|^2|Y^{\mu}_b|^2|Y^{\mu}_c|^2|Y^{\mu}_d|^2\\
&\hskip3,5cm+\sum_{I_3}H_{\mathbf p, \mathbf q}\bar{H}_{\mathbf p', \mathbf q'}(|Y^{\mu}_i|^4|Y^{\mu}_j|^2|Y^{\mu}_k|^2+Y^{\mu}_i|^3|Y^{\mu}_j|^3|Y^{\mu}_k|^2)\\
&\hskip3,5cm+\sum_{I_4}H_{\mathbf p, \mathbf q}\bar{H}_{\mathbf p', \mathbf q'}(|Y^{\mu}_i|^4|Y^{\mu}_j|^4+|Y^{\mu}_i|^6|Y^{\mu}_j|^2+|Y^{\mu}_i|^5|Y^{\mu}_j|^3)\Big\}\\
&\le \tilde{c}\Big(\sum_{p_1, p'_1, p_2, q_2}(H+\Delta H+H\Delta+\Delta H\Delta)_{p_1 p_2, q_1 q_2}
(\bar{H}+\Delta \bar{H}+\bar{H}\Delta+\Delta \bar{H}\Delta)_{p'_1 p_2,p'_1 q_2}\\
&\hskip2cm+\tr (H+\Delta H+H\Delta+\Delta H\Delta)(H+\Delta H+H\Delta+\Delta H\Delta)^*
\\
&\hskip9cm
+|I_3|c^2 n \tau^{2}+|I_4|c^2 n^2 \tau^4\Big).
\end{align*}
Since $\Delta^2=I$ and $|I_3|=c_1n^3$, $|I_2|=c_2n^2$ we have:
\begin{align*}
R_1\le \tilde{c}\Big(\sum_{p_1, p'_1, p_2, q_2}C_{p_1 p_2, q_1 q_2}C^*_{p'_1 p_2,p'_1 q_2}+\tr HH^*+\tr \Delta HH^*+c n^4 \tau\Big),
\end{align*}
where
\begin{align*}
C=H+\Delta H+H\Delta+\Delta H\Delta.
\end{align*}
Denote by $\tilde{C}$ an $n\times n$ matrix with coordinates
\begin{align*}
\tilde{C}_{p_2q_2}=\sum_{p_1=1}^{n}C_{p_1p_2,p_1q_2}.
\end{align*}
Then
\begin{align*}
R_1\le c\Big(\tr \tilde{C}\tilde{C}^*+\tr HH^*+\tr \Delta HH^*+cn^4\tau\Big).
\end{align*}
It is easy to see that $|\tilde{C}|<n|H|<nc$, hence
\begin{align*}
R_1\le c(n^3+n^2+n^4\tau).
\end{align*}
Divide the set $\{(\mathbf{p}, \mathbf{p}')\}$ of all possible indexes into four sets $\{I_i\}_{i=1}^4$ such that $(\mathbf{p}, \mathbf{p}')\in I_i$ if there are exactly $i$ different numbers in the set $(p_1, p_2, p'_1, p'_2)$.
The matrices $H$ and $J$ are bounded, so in view of (\ref{cond_1}) and (\ref{cond_2})
\begin{align*}
&\hskip2cmR_2\le c\mathbf{E}\Big\{\sum_{I_1}|Y^{\mu}_1|^8+\sum_{I_2}(|Y^{\mu}_1|^4|Y^{\mu}_2|^4+|Y^{\mu}_1|^6|Y^{\mu}_2|^2)+\sum_{I_3}|Y^{\mu}_1|^4|Y^{\mu}_2|^2|Y^{\mu}_3|^2\\
&+\sum_{I_4}(|Y^{\mu}_1|^2|Y^{\mu}_2|^2-1)(|Y^{\mu}_3|^2|Y^{\mu}_4|^2-1)\Big\}=c(|I_1|n^3\tau^6+|I_2|n^2\tau^4+|I_3|n\tau^2+|I_4|o(1))\\
&\hskip11cm=cn^4(\tau+o(1)).
\end{align*}
Note that if the set of indexes $\{p_1, p_2, p'_1, p'_2, q'_1, q'_2\}$ has more than 3 or less than 2 different numbers then
\begin{align*}
\mathbf{E}\Big\{\Big(|Y^\mu_{p_1}|^2|Y^\mu_{p_2}|^2-1\Big)\bar{Y}^\mu_{p'_1}\bar{Y}^\mu_{p'_2} Y^\mu_{q'_1}Y^\mu_{q'_2}\Big\}=0.
\end{align*}
Other terms we divide into sets $I_1$ (3 different numbers) and $I_2$ (2 different numbers). Similarly to previous case
\begin{align*}
R_3\le c\Big (\sum_{I_1}n\tau^2+\sum_{I_2}n^2\tau^4\Big )=cn^4\tau.
\end{align*}
At last, we get:
\begin{align*}
\mathbf E_{\mu}\{(N^{-1}r^\mu)^2\}\le o(1)+c\tau.
\end{align*}
Since this inequality is true for every $\tau$, we have
\begin{align*}
\mathbf E_{\mu}\{(N^{-1}r^\mu)^2\}= o(1).
\end{align*}

\vspace{0,5cm}

$(ii)$
According to (\ref{proek}),
\begin{align*}
(F(G-G^\mu))_{\mathbf i, \mathbf j}=-\dfrac{N^{-1}(FG^\mu
X^\mu)_{\mathbf i} \overline{(G^\mu X^\mu)}_{\mathbf
j}}{1+N^{-1}(G^\mu X^\mu, X^\mu)}.
\end{align*}
Hence
\begin{align*}
\left|\tr(F(G-G^\mu))\right|=\left|\dfrac{N^{-1}(FG^\mu X^\mu,
G^\mu X^\mu)}{1+N^{-1}(G^\mu X^\mu, X^\mu)}\right|\le
\dfrac{|F|\left|((G^\mu)^*G^\mu X^\mu, X^\mu)\right|}{\left|\Im
(G^\mu X^\mu, X^\mu)\right|}.
\end{align*}
From the other hands by the spectral theorem
\begin{align*}
(G^\mu X^\mu, X^\mu)=\sum_{k=1}^{m-1}\dfrac{(v^k,
X^\mu)^2}{\lambda _k-z},
\end{align*}
where $\{\lambda_k\}$ are eigenvalues of $G^\mu$ and $\{v^k\}$ are
eigenvectors of $G^\mu$. Then
\begin{align*}
\left|\Im(G^\mu X^\mu, X^\mu)\right|=|\Im
z|\sum_{k=1}^{m-1}\dfrac{|(v^k, X^\mu)|^2}{(\lambda
_k-z)(\lambda^* _k-z)}.
\end{align*}
Besides,
\begin{align*}
((G^\mu)^*G^\mu X^\mu, X^\mu)=\sum_{k=1}^{m-1}\dfrac{|(v^k,
X^\mu)|^2}{(\lambda _k-z)(\lambda^* _k-z)}.
\end{align*}
Finally we get
\begin{align*}
\dfrac{1}{N}\tr F(G-G^\mu)\le\dfrac{|F|}{N|\Im z|}=O(N).
\end{align*}


\vspace{0,5cm}

$(iii)$
To prove the lemma we need the follow statement of martingale
bounds (see e.g.~\cite{G:01} for results and references):
\begin{lemma}\label{var}
Let $\{Y^\mu\}_{\mu=1}^m$ be a sequence of i.i.d random vectors of
$\mathbb R^n (\mathbb C^n)$. Assume that the function
$\phi:\mathbb R^{nm} (\mathbb C^{nm})\to \mathbb C$ is a bounded
Boreal function such that
\begin{align*}
\sup_{X^1,\ldots,X^\mu\in \mathbb R^n(\mathbb
C^n)}|\phi-\phi^\mu|\le c,
\end{align*}
where $\phi^\mu=\phi\mid_{X^\mu=0}$. Then
\begin{align*}
\var\{\phi(Y^1,\ldots,Y^\mu)\}\le 4c^2 m.
\end{align*}
\end{lemma}

Take $\phi=\tr (FG)$. Then, using representation (\ref{proek}), we
obtain
\begin{align*}
|\phi-\phi^\mu|=|\tr G-\tr G^\mu|=\left|\dfrac{N^{-1}(G^\mu FG^\mu
X^\mu, X^\mu)}{1+N^{-1}(G^\mu X^\mu, X^\mu)}\right|.
\end{align*}
Similarly to the proof of the previous result we have
\begin{align*}
\left|\dfrac{N^{-1}(G^\mu FG^\mu X^\mu, X^\mu)}{1+N^{-1}(G^\mu
X^\mu, X^\mu)}\right|\le c|\Im z|^{-1}.
\end{align*}
Thus,
\begin{align*}
|\phi-\phi^\mu|\le c|\Im z|^{-1}.
\end{align*}
So, according to Lemma \ref{var},
\begin{align*}
\var\{g_n\}\le 4c^2 c_n/N.
\end{align*}

$\square$

\textbf{Acknowledgements.} The author is grateful to Prof.L.A.Pastur for statement of the
problem and fruitful discussion.

\end{document}